\documentclass[twocolumn]{aastex631}

\begin{document}

\title{Variations in the 6.2~$\mu$m polycyclic aromatic hydrocarbon band in Active Galactic Nuclei- and Starburst-dominated galaxies}

\correspondingauthor{Carla M. Canelo}
\email{camcanelo@gmail.com}

\author[0000-0002-7720-0858]{Carla M. Canelo}
\affiliation{Instituto de Matem\'atica, Estat\'istica e F\'isica, Universidade Federal do Rio Grande, \\ 96201-900 Rio Grande do Sul, Brazil}

\author[0000-0002-3496-5711]{Dinalva A. Sales}
\affiliation{Instituto de Matem\'atica, Estat\'istica e F\'isica, Universidade Federal do Rio Grande, \\ 96201-900 Rio Grande do Sul, Brazil}

\author[0000-0002-7512-6033]{Vitor Avelaneda}
\affiliation{Instituto de Matem\'atica, Estat\'istica e F\'isica, Universidade Federal do Rio Grande, \\ 96201-900 Rio Grande do Sul, Brazil}

\author[0000-0003-0306-0028]{Alexander G.G.M. Tielens}
\affiliation{Leiden Observatory, Leiden University, Einsteinweg 55, 2333 CC Leiden, The Netherlands}

\author[0000-0002-0718-5103]{Miriani Pastoriza}
\affiliation{Departamento de Astronomia, Instituto de F\'isica, Universidade Federal do Rio Grande do Sul, \\ Porto Alegre, 91501-970, Rio Grande do Sul, Brazil }

\author[0009-0007-8204-1234]{Am\^ancio C. S. Fria\c{c}a}
\affiliation{Departamento de Astronomia, Instituto de Astronomia, Geof\'isica e Ci\^encias Atmosf\'ericas, Universidade de S\~ao Paulo, \\ S\~ao Paulo, 05508-090, S\~ao Paulo, Brazil}



\begin{abstract}

Polycyclic aromatic hydrocarbons (PAHs) are fundamental to  understanding  the interstellar medium (ISM) of several astrophysical objects. Normally present in Starburst (SB) galaxies, they have also been more frequently detected in active galaxy nuclei (AGNs), suggesting an inner dusty torus that can shield the radiation from the central black role. In this work, we analyze the 6.2~$\mu$m PAH band of SB-, AGN- and mixed-dominated spectra from 175 IDEOS database galaxies. After fitting of the band, the sources were distributed into the Peeters' A, B and C classes according to their profile peak positions. Class A objects are predominant in 80\% of the entire sample, which could indicate the presence of PAHs with nitrogen incorporation. The water ice absorption at 6.0~$\mu$m was also studied in eleven objects, and it affected the PAH band poorly. A prominent second spectral feature after 6.3~$\mu$m is  present in ten galaxies. Fitting both PAH profiles at 6.2~$\mu$m changes all the fit results: the first profile is consistently blue-shifted and classified as class A due to the presence of the second component.  Further studies are needed to better comprehend these PAH trends in galactic environments.

\end{abstract}

\keywords{Extragalactic astronomy --- Infrared galaxies --- Infrared spectroscopy ---Interstellar molecules --- Astrochemistry -- Starburst galaxies -- Active galactic nuclei}


\section{Introduction} \label{sec:intro}

Polycyclic Aromatic Hydrocarbons (PAHs) are indispensable molecules when  considering the class of macromolecules and their relevance in the study of the Universe. Due to their effective arrangement, they can accumulate about 10 to 20\% of  carbon in the interstellar medium (ISM), being the dominant organic material in space \citep{Eh06}. Those molecules are also fundamental for studying the dust emission of astrophysical sources, once the mid-infrared (MIR) spectra of galactic and extragalactic objects are dominated by strong emission features at 3.3, 6.2, 7.7, 8.6, 11.3 and 12.7$\mu$m, generally attributed to PAHs and related species \citep{Li04}.

Recent works have explored in more detail the observation and modeling of PAHs in local sources. \citet{Sidhu2022}, for instance, presented a charge distribution based model to compute the IR spectrum of these molecules using recent measurements or quantum chemical calculations of specific PAHs, that can be applied to photodissociation regions (PDRs). In addition, \citet{Knight2022} used observations from the Stratospheric Observatory for Infrared Astronomy (SOFIA\footnote{\url{https://www.sofia.usra.edu/}}), Herschel Space Observatory\footnote{\url{https://www.herschel.caltech.edu/}} and Spitzer Space Telescope\footnote{\url{https://www.spitzer.caltech.edu/}} together with PDR modeling to characterize spatial variations of PAH emission in the Reflection Nebula NGC~1333. They concluded that the 6–9 $\mu$m PAH bands arise from multiple PAH sub-populations with different underlying molecular properties. However, although PAHs are abundant and ubiquitous in almost all astrophysical environments related to gas, dust and ultraviolet (UV) photon illumination \citep{Tielens08}, the study of PAHs has focused mainly on galactic objects. 

In an extragalactic point of view,  PAH features are often present in star-forming systems, diminished and modified in high-intensity starbursts, and eventually disappear in active galactic nuclei (AGN) systems \citep{yan07}. More specifically, starburst spectra are dominated by strong emission of these features, not only in the continuum shape \citep{Genzel00} but also in the 5 -- 8~$\mu$m spectral range with the 6.2~$\mu$m band and the blue wing of the 7.7~$\mu$m PAH complex \citep{Brandl06} as well. Furthermore, starburst galaxies and most ULIRGs \citep[Ultra-Luminous Infrared Galaxies, ][]{yan05} can also present aromatic, dust grain and absorption features in their MIR spectra. Observations of the M82 starburst from James Webb Space Telescope (JWST\footnote{\url{https://webb.nasa.gov/}, \url{https://esawebb.org/}}) revealed prominent plumes of PAH emission extending outward from the central starburst region, together with a network of complex filamentary substructures and edge-brightened bubble-like features \citep{Bolatto2024}. Regarding PAHs in M86's outflows, suggest that the extreme conditions translate into CO not tracing the full budget of molecular gas in smaller clouds, perhaps as a consequence of photoionization and/or emission suppression of CO molecules due to hard radiation fields from the central starburst \citep{Villanueva2025}.

In the case of active galaxy nuclei (AGNs), for a long time it was believed that such galaxies that have supermassive black holes (SMBH), have an intense and energetic radiation field from the SMBH accretion disk capable of destroying the molecules of PAHs close to the black holes. The lack of data with high angular and spectral resolution helped to make it difficult to understand the physical-chemical properties of PAHs and their use as a diagnostic tool for the ionization source in galaxies. Nevertheless,  the spectra obtained with the ISO\footnote{\url{https://www.cosmos.esa.int/web/iso}} and Spitzer space telescopes  greatly expanded the analyses of PAHs in extragalactic sources, allowing the production of diagnostic diagrams using the band intensities of  PAHs as a tool to distinguish contributions coming from different types of ionization sources \citep[e.g.][]{genzel1998,lutz1998,rigopoulou1999,tran2001,Smith07,galliano2006,Galliano08,gordon2008,sales10,sales2011,sales13,sales2015}. Using multi-wavelength observations, \citet{Shimizu2019} investigated AGN feedback in NGC 5728, revealing an ionized gas outflow driven by the AGN that reaches 250 pc with a mass outflow rate of 0.08 M$_{\odot}$ yr$^{-1}$. Although potential signatures of molecular gas removal are observed, the circumnuclear ring structure and overall gas kinematics remain largely unaffected by AGN feedback.

Such diagnostic diagrams showed that the MIR emission of AGNs is clearly different from those dominated by a Starburst nucleus -- galaxy dominated by star formation
\citep[][for example]{lutz1998,genzel1998,laurent2000,gordon2008,sales10}. From these works it is possible to notice that the spectrum of an extremely luminous AGN presents the PAH bands very weak or even absent, once the radiation coming from the active nucleus would be destroying the dust located close to the nucleus. On the other hand, when analyzing the spectra of Starbursts, these bands are very intense, and a lot of effort has been given in order to discriminate whether the radiation that excites these molecules comes from the stars or the AGN. Currently, the detection of PAHs at up to a dozen parsecs from the AGN indicates the presence of a dusty material, such as a nuclear torus or disk, which allows these molecules to survive in environments so close to AGNs \citep{sales10,sales13,ruschel-dutra14, Alonso-Herrero14, Alonso-Herrero16, Monfredini19}. Furthermore, \citet{Salvestrini2022} corroborate with such studies showing that molecular gas content only is reduced in regions of sub-kpc size, where the emission from the accreting SMBH dominates. In spite of that, they also show that the impact of AGN activity on the ISM is clearly visible as suppression of the PAH luminosity in the MIR.

As can be seen, the first JWST results are beginning to expand our knowledge of PAHs with its unprecedented resolution. \citet{Chastenet2023} were able to infer changes in the average PAH size and ionization state across the different galaxy environments in the nearby galaxies  NGC 628, NGC 1365, and NGC 7496. They found evidence of heating and/or changes in PAH size in regions with higher molecular gas content as well as increased ionization in regions with higher H$\alpha$ intensity. \citet{Leroy2023} observed the relationship between PAH and CO rotational line emission from massive star-forming galaxies, consistent with PAH emission arising not just from CO-bright gas but also from atomic or CO-dark gas. On the other hand, \citet{Sutter2024} found that the PAH fraction steeply decreases in H II regions, revealing the destruction of these small grains in regions of ionized gas.  These studies exemplify the importance of exploring the characteristics of PAHs in more detail.

\subsection{The Peeters' classification}

A more in-depth analysis of PAH observations in different astrophysical environments has revealed variations in their emission profiles. In general, these observational spectral variations can be divided into variations in the relative intensities of different mid-infrared (MIR) emission bands and variations in the peak position and profile of these bands \citep{Tielens2008}. To better understand these variations, \citet{Peeters02} suggested a systematic classification of PAH bands at 6.2, 7.7, and 11.2~$\mu$m into three classes --- A, B, and C --- depending on the interpretation of the peak position variation of the profiles. Subsequently, this classification was extended to other bands by \citet{died04}.

The Peeters' classification have been applied to several galactic objects and indicated a correlation of the physical and chemical conditions of the sources with the PAH profiles \citep[e.g.][]{died04, Candian15, Peeters17, Shannon19}.  Typically, class A objects are connected to interstellar material illuminated by stars, such as H\,II regions, reflection nebulae, and the interstellar medium in general (in the Galaxy and in extragalactic objects). Class B, on the other hand, is more closely related to circumstellar material, planetary nebulae, various post-AGB objects, and Herbig Ae/Be stars. Finally, class C has been attributed to a few extremely carbon-enriched post-AGB objects \citep{Peeters02,Tielens2008}. 

The same approach was first performed  on an extragalactic sample by \citet{Canelo18} and \citet{Canelo21b}. Specifically considering the 6.2~$\mu$m PAH band, \citet{Canelo18} demonstrated a domination of the blueshift of the central wavelength of the band, a characteristic of class A profiles \citep{Peeters02}, in galaxies dominated by starburst emission. The case of the 6.2~$\mu$m band rises an astrobiological interest because the observed blueshift of the band in astrophysical objects has been well reproduced by the incorporation of nitrogen on inner rings of PAHs \citep[Polycyclic Aromatic Nitrogen Heterocycles (PANHs),][]{Hud05}.   In fact, the recent theoretical study of infrared spectra of interstellar PAH molecules with N, NH, and NH$_{2}$ incorporation  using density functional theory (DFT) by  \citet{Vats2022} suggests that strong bands at 6.2 and 11.2~$\mu$m can arise from the same charge state of some N-containing PAHs, arguing that there might be some N-abundant astronomical regions where these bands ratio is not a direct indicator of the PAHs' ionization. 

\citet{Ricca2021} also revised this issue considering the quantum chemical spectra collected from NASA Ames PAH IR Spectroscopic Database\footnote{\url{https://www.astrochemistry.org/pahdb/}} \citep[PAHdb;][]{Boersma2014B, Bauschlicher2018, Mattioda2020} and polarization corrections that resulted in blue shifts in the band position. However, they also estimated an upper limit of 12~\% in the derived abundances of PANHs in class A spectra. This could indicate the presence of PANHs and their important contribution to the MIR emission in the ISM of these sources. Despite the issue in the PAHN identification, which is not addressed here, we cannot ignore a potential source of nitrogen in AGN and starburst environments. 

In this sense, here we refine and expand the statistical analysis of the 6.2~$\mu$m PAH band of \citet{Canelo18} by improving their sample with more accurate data. {Our main goal is to use Peeters' classification and its relationship with the physicochemical properties of astrophysical environments to compare and distinguish AGN from starburst galaxies, as well as reveal the first insights on the presence of PANHs in AGNs.} In addition, together with further analysis of the 7.7, 8.6 and 11.2~$\mu$m PAH band, we aim to improve the Salles diagnostic diagram \citep[e.g.][]{sales10,sales13,ruschel-dutra14}, which can distinguish AGN from starburst objects. Finally, our study is extremely important as a future approach that can be improved with the high sensitivity of the James Webb Space Telescope\footnote{\url{https://webb.nasa.gov/}}. This paper is structured as follows: Section \ref{sec:data} presents the selection of our sample  and the data analysis performed in the spectroscopic data is described in Section \ref{sec:analysis}. The results are discussed in Section \ref{sec:results} and Section \ref{sec:conclusion} presents the summary and conclusion. A brief comparison between the analysis of \citet{Canelo18} and ours is shown in Appendix~\ref{sec:ap}.

\section{Data Selection}
\label{sec:data}

The analysis of the 6.2~$\mu$m band of \citet{Canelo18} revealed a predominance of class A objects in 67\% of their SB galaxies, which could be related with the starburst environment. In order to expand their analysis and compare the variations in this PAH profile according to the different astrophysical properties of the galaxies, we expanded our sample with other SB and AGN-dominated objects with more accurate data. 

While \citet{Canelo18} used the ATLAS project database \citep{caballero} -- which contains reduced spectra from the Infrared Spectrograph \citep[IRS, ][]{Houck04} of the Spitzer Space Telescope \citep{Werner04} extracted from PostScript figures uploaded to arxiv.org -- we used the spectra from Infrared Database of Extragalactic Observables from Spitzer\footnote{\url{http://ideos.astro.cornell.edu/}}  \citep[IDEOS,][]{Spoon_2022}.  The IDEOS database provides mid-infrared diagnostics and spectra for more than 3,330 galaxies/galactic nuclei from spectroscopic measurements obtained by the IRS on Spitzer. The low-resolution spectra in staring-mode have been drawn and vetted from the  Cornell Atlas of Spitzer/Infrared Spectrograph Sources \citep[CASSIS,][]{Lebouteiller2011}. The redshifts of the galaxies were also taken from 
NASA/IPAC Extragalactic Database\footnote{\url{https://ned.ipac.caltech.edu/}}  (NED) or from literature.

The classification of the sources into  AGN- and SB-dominated galaxies  was based on a diagnostic diagram that combined the ice-corrected equivalent width of the 6.2~$\mu$m PAH feature with the silicate strength into a grid of 3 × 3 mid-IR classes, from 1A to 3C.  \citep[see][for more details]{Spoon_2022}. We then divide our sample in four types (hearafter, grouped IDEOS classification): SB, corresponding to class 1C PAH-dominated spectra typical for starburst galaxies; AGN~1, corresponding to class 1A objects with a hot-dust-dominated spectrum typical for AGNs; AGN~2, corresponding to class 3A galaxies with an absorption-dominated spectra typical for centrally heated dust shells or tori; and, finally, the mixed-dominated objects are the galaxies in between these classes. We selected the objects with highest signal-to-noise ratios (S/N $\geq$ 10 at 6.6m$\mu$m), focusing meanly near the 6.2~$\mu$m region to not compromise our analysis, resulting in a total of 175 galaxies.  Their IDEOS information and the new grouped classification is presented in Table~\ref{tab:info}.  {In particular, our selected sample does not contain AGN~2 objects and is composed of 14.3\% of SB, 34.3\% of AGN~1, and 51.4\% of mixed-dominated galaxies. From them, 75 objects are also present in \citet{Canelo18}. When applying the new grouped IDEOS classification to the previous sample from \citet{Canelo18}, we find a notably different composition: 70\% of mixed-dominated galaxies, 29\% of SB, and only 1\% of AGN~1. This comparison highlights the improved classification accuracy and better separation of galaxy types achieved with our refined methodology.}

\begin{deluxetable*}{ccccccccc} 

\tablewidth{0pt}
\tablecaption{IDEOS Information of the sample. \label{tab:info}}
\tablehead{
\colhead{IDEOS\_name} & \colhead{z} & \colhead{MidIRClass} & \colhead{PAH$_{6.2}$EW} & \colhead{e\_PAH$_{6.2}$EW} & \colhead{Sil$_{Strength}$} & \colhead{e\_Sil$_{Strength}$} & \colhead{grouped\_IDEOS\_class} \\  
\colhead{} & \colhead{} & \colhead{} & \colhead{} & \colhead{$\mu$m} & \colhead{$\mu$m} & \colhead{} & \colhead{}  
}
\startdata
00000001\_0 & 0.018 & 3B & 0.273 & 0.001 & -2.736 & 0.004 & mixed \\
00000002\_0 & 0.042 & 1A & 0.021 & 0.001 & -0.630 & 0.002 & AGN1 \\
00000004\_0 & 0.149 & 1A & 0.011 & 0.001 & 0.081 & 0.003 & AGN1 \\
00000007\_0 & 0.241 & 1A & 0.008 & 0.000 & 0.565 & 0.011 & AGN1 \\
00000010\_0 & 0.158 & 1A & 0.003 & 0.000 & 0.069 & 0.002 & AGN1 \\
10103296\_0 & 0.1292 & 2B & 0.457 & 0.0368 & -1.297 & 0.121 & mixed \\
10103552\_0 & 0.129 & 2C & 0.6025 & 0.0227 & -0.917 & 0.055 & mixed \\
10103808\_0 & 0.11746 & 1B & 0.2813 & 0.0109 & -0.421 & 0.023 & mixed \\
10104320\_0 & 0.14511 & 1C & 0.5238 & 0.0218 & -0.682 & 0.05 & SB \\
10105344\_0 & 0.09206 & 2C & 0.5415 & 0.0145 & -1.254 & 0.037 & mixed
\enddata
\tablecomments{Table 1 is published in its entirety in the electronic edition of the {\it Astrophysical Journal}.  A portion is shown here for guidance regarding its form and content. {Label descriptions: IDEOS\_name is the name in the IDEOS catalog, z is the redshift, MidIRClass is the Mid-IR classification based on IDEOS, PAH$_{6.2}$EW is the equivalent width of the 6.2~$\mu$m PAH band, e\_PAH$_{6.2}$EW is the error in equivalent width of the 6.2~$\mu$m PAH band, Sil$_{Strength}$ is the silicate strength at 9.8~$\mu$m, e\_Sil$_{Strength}$ is the error in silicate strength, and grouped\_IDEOS\_class is the grouped IDEOS classification based on MidIRClass.}}
\end{deluxetable*}

\section{Data analysis}
\label{sec:analysis}

\subsection{Continuum subtraction}
\label{sec:continuum}

In order to adequately fit and analyze the 6.2~$\mu$m PAH feature, the underlying continuum emission must be subtracted from the spectra. From the different forms available to stipulate the continuum in the literature, one approach is to fit a spline to the spectra, {which is a smooth curve constructed from segments of polynomials that can be shaped by a set of anchor points} \citep[e.g.][]{Brandl06, Galliano08, Peeters17, Canelo18, Canelo21b}. According to \citet{Peeters17}, there are two distinct spline methods based on their anchor points used in the fitting. In general, the global spline is determined by anchor points at roughly 5.4, 5.8, 6.6, 7.2, 9.0, 9.3, 9.9, 10.2, 10.9, 11.7, 12.1, 13.1, 13.9, 14.7 and 15.0~$\mu$m. The local spline, on the other hand, includes another anchor point near 8.2~$\mu$m, considering the broad emission feature at this wavelength as part of the continuum emission. 

Although the fit of the 6.2~$\mu$m is independent of the spline decomposition method \citep[e.g.][]{Smith07, Galliano08}, the use of local or global spline does influence the 7.7~$\mu$m complex intensities \citep[e.g.][]{Peeters17, Canelo21b}. {In addition, other codes such as \textsc{PAHFIT} \citep{Smith07}, its variation using the Markov Chain Monte Carlo method \textsc{PAHFIT-MCMC} \citep{sales2015,Gleisinger2020}, and \textsc{CAFE} \citep{Marshall2007, Diaz-Santos2025}, could also be used to decompose the continuum. However, normally they do not fit well the PAH plateau emission in the 5--18~$\mu$m range. Instead, the plateau is diluted in the PAH features when fitted by Drude profiles \citep[e.g.][]{Smith07, Spoon_2022}. We included an example of this effect in Appendix-\ref{sec:ap2}.  To avoid contamination of the plateau, we used the spline method for the continuum and Gaussian profiles for the PAH bands.}

We also allowed a small variation of 0.05 -- 0.1~$\mu$m in the anchor points positions to choose the minimum values and avoid unwanted removal of PAH flux. Although small variations in the anchor points may  induce a systematic error up to 20\% in the 6.2~$\mu$m band amplitude  \citep{Canelo21b},  its central wavelength and FWHM (full width at half maximum) {remain} practically the same with less than 1\% variation and, therefore, the Peeters' classification is not influenced by potential spline uncertainties. Our galaxies may present different extinction contributions or even low values. The extinction at longer wavelengths is dominated by the silicate features, specially at  9.7~$\mu$m, which could be an emission or absorption feature {\citep[e.g.][]{Beriao2008}}. Consequently, we have not performed any previous extinction correction and neither used objects with higher absorption levels. Despite this, considering the extinction curve of \citet{Hensley2020}, the results of \citet{Canelo21b} suggest that Peeters' classes are little or quite not influenced by the extinction for the 6.2~$\mu$m band.

There is no evidence of strong H2 S(6) emission at 6.1~$\mu$m and, therefore, it was not considered. For a more detailed decomposition model considering ionic emission and other spectral components, one may use the PAHFIT model decomposition of \citet{Smith07}. However, this code does not allow for the variation of the band emission central wavelengths and, therefore, is not indicated in our work. Nevertheless, the PAHFIT decomposition of the continuum, ionic and molecular emission lines and silicate absorption has already been used to fit the continuum of SB-dominated galaxies in \citet{Canelo18}, before the fitting of the 6.2~$\mu$m band. They also performed a comparison of the band fitting considering the continuum subtracted by the PAHFIT model and by a spline. Both methods showed the same results for the 6.2~$\mu$m band.

\subsection{Gaussian fit of the band}
\label{sub-analysis}

We applied the same fitting method used by \citet{Canelo18}. Their \textsc{python}-based script estimates the central wavelength, amplitude, and FWHM of this feature through the optimization algorithms from the submodule \textit{scipy.optmize.curve\_fit}. This band profile is well-known for its asymmetry \citep[e.g.][]{Peeters02,Tielens08}, especially in the case of class A objects, in which this feature presents a sharp blue rise and a red tail with central wavelength varying up to 6.23~$\mu$m \citep{Peeters02}. This asymmetry tends to decrease in objects of class B and C, and the peak position could be shifted to wavelengths {longer} than  6.29~$\mu$m in the last scenario. During the fitting process, this asymmetry can produce a deviation of the central wavelength to the redder values.  Although asymmetry is not an major factor in determining the profile itself, it is worth mentioning that one possible contributor to this red tail is an-harmonic hot-band emission which is a natural consequence of the PAH model \citep{hudgins99}.

This band is generally fitted with just one profile, in spite of which kind (Gaussian, Lorentz or Drude, for instance). However, \citet{Canelo18} showed that this asymmetric band could be better fitted with the addition of a second profile rather than to reduce the fit range for some objects.  In these cases, both profiles are fitted together by summing two Gaussian profiles and using the same optimization algorithms. 

Moreover, the water ice absorption at 6~$\mu$m could compromise the 6.2~$\mu$m flux band. In order to approach this issue, we include another Gaussian profile in the fitting. \citet{Spoon_2022} have also fitted this feature with a Gaussian profile, as well as the aliphatic CH deformation absorption band at 6.85~$\mu$m. According to their model, only the continuum component is subject to attenuation by water ice and/or aliphatic hydrocarbons, because these emission lines and PAH features are assumed to have a circumnuclear origin and are unaffected by water-ice and aliphatic hydrocarbon absorption occurring within the nucleus in models for buried galactic nuclei \citep[e.g.][]{Veilleux09}. In fact, the CH absorption band (i.e. the 6,8~$\mu$m CH$_2$/CH$_3$ deformation modes) has not shown any interference with the fit of the 6.2~$\mu$m band, even when considering two profiles. 

With the Gaussian fit results, it is possible to group the galaxies into the Peeters' classes. In the case of the 6.2~$\mu$m band, the classification system is simply based  on the divergence of its central wavelength ({$\lambda_c$}): class A corresponds to $\lambda_c \leq$~6.23~$\mu$m, class B to 6.23~$\mu$m~\textless~$\lambda_c \leq$~6.29~$\mu$m, and class C to $\lambda_c$~\textgreater~6.29~$\mu$m.

\section{Results and Discussion}
\label{sec:results}

\begin{figure}[ht!]

\centering
\includegraphics[width=\linewidth, keepaspectratio]{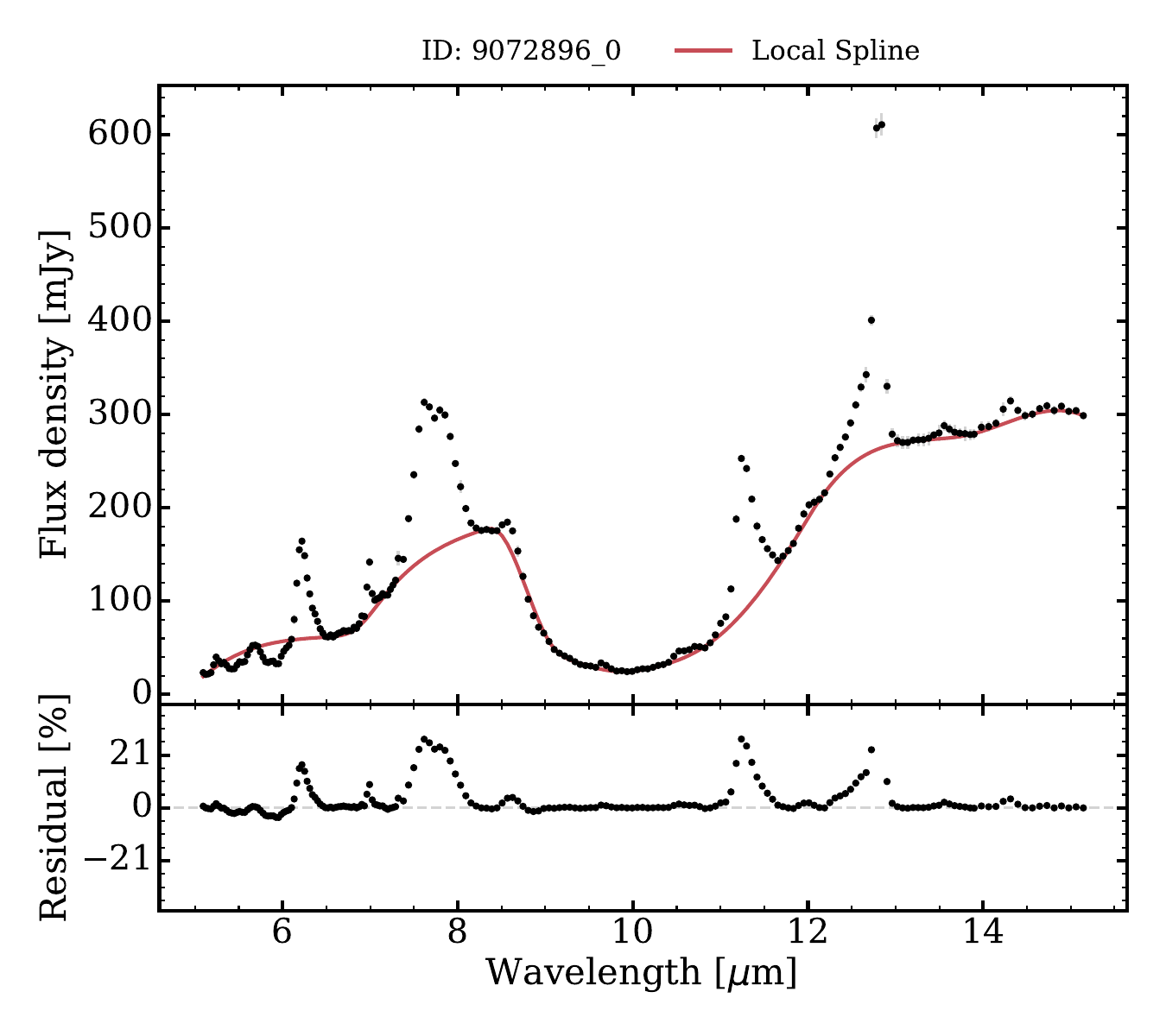}
\caption{Example of the fits performed in the sample for  Spline decomposition.  The complete figure set (175 images) is available in the online journal. }
\label{fig:spl}
\end{figure}

\begin{figure}[ht!]

\centering
\includegraphics[width=\linewidth, keepaspectratio]{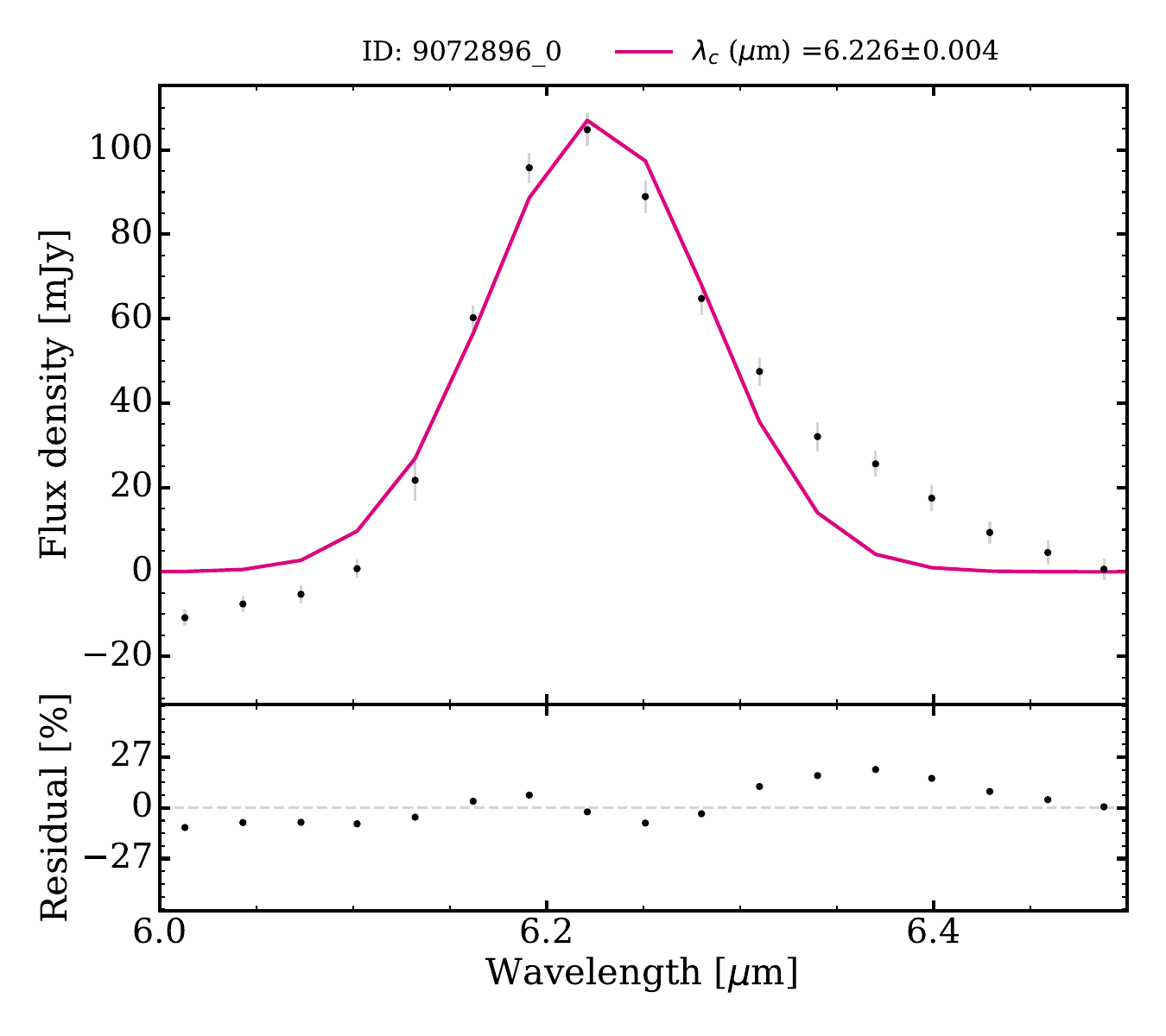}
\caption{Example of the fit results of one PAH band at 6.2~$\mu$m.  The complete figure set (175 images) is available in the online journal. }
\label{fig:1p}
\end{figure}

\begin{figure}[ht!]

\centering
\includegraphics[width=\linewidth, keepaspectratio]{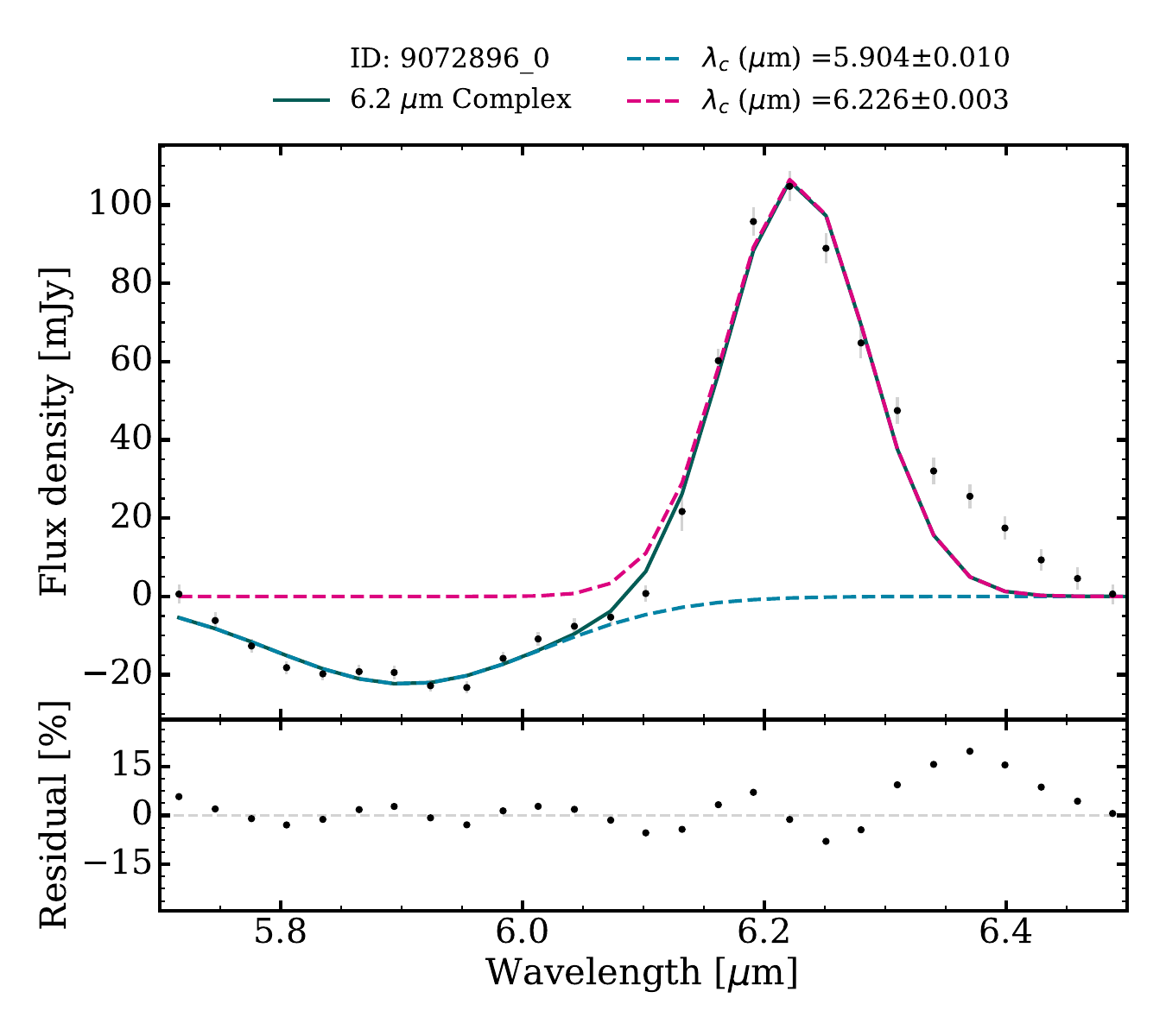}
\caption{Example of the fit results of water ice absorption at 6.0~$\mu$m and PAH band at 6.2~$\mu$m.  The complete figure set (11 images) is available in the online journal. }
\label{fig:1p-ice}
\end{figure}

\begin{figure}[ht!]

\centering
\includegraphics[width=\linewidth, keepaspectratio]{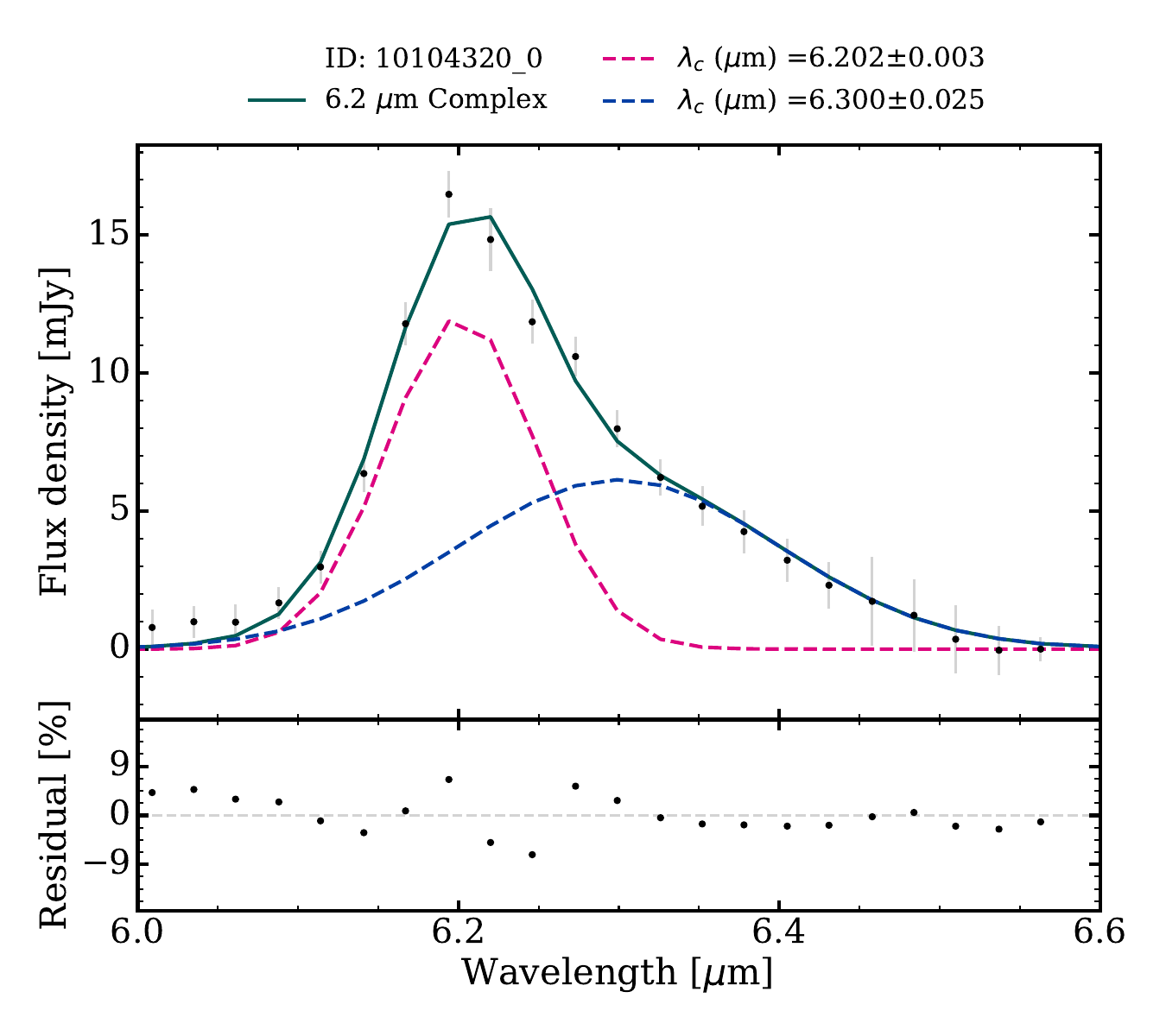}
\caption{Example of the fit results of two PAH bands at 6.2 and 6.3~$\mu$m.  The complete figure set (10 images) is available in the online journal. }
\label{fig:2pah}
\end{figure}

The results are presented in Table~\ref{tab:fits}, where Amp is the amplitude, Wl\_c is the central wavelength and FWHM is the full width at half maximum, together with their derived uncertainties.  The Band and fit\_type columns represent the type of fit applied and the resulting features, with "1p" as the single PAH profile ("Band" as 6.2), "2p" as double PAH profiles ("Band" as 6.2 and 6.4), and "ice" as the single PAH profile with ice profile ("Band" as 6.0 and 6.2). Naturally, the Peeters' classes are only attributed to the 6.2~$\mu$m PAH band. The spline and band fits are exemplified in Figures~\ref{fig:spl} to ~\ref{fig:2pah}, but all plots are available online. The data points are represented by the dots with the vertical error-bars as uncertainties.

\begin{deluxetable*}{cccccccccc}

\tablewidth{0pt}
\tablecaption{Best-fit results for the 6.2~$\mu$m band and its Peeters' classification, together with results for 6.0 and 6.4 features when available.   \label{tab:fits}}
\tablehead{
\colhead{IDEOS\_name} & \colhead{Band} & \colhead{Amp} & \colhead{e\_Amp} & \colhead{$\lambda_c$} & \colhead{e\_$\lambda_c$} & \colhead{FWHM} & \colhead{e\_FWHM} & \colhead{fit\_type} & \colhead{Peeters\_class} \\
\colhead{} & \colhead{} & \colhead{mJy} & \colhead{mJy} & \colhead{$\mu$m} & \colhead{$\mu$m} & \colhead{} & \colhead{} & \colhead{} & \colhead{}
}
\startdata
00000001\_0 & 6.0 & -49.293 & 1.919 & 5.930 & 0.008 & 0.293 & 0.022 & ice & --- \\
00000001\_0 & 6.2 & 88.574 & 7.466 & 6.224 & 0.005 & 0.120 & 0.011 & 1p & A \\
00000001\_0 & 6.2 & 90.919 & 4.328 & 6.219 & 0.003 & 0.130 & 0.007 & ice & A \\
00000002\_0 & 6.2 & 78.137 & 1.496 & 6.209 & 0.002 & 0.172 & 0.007 & 1p & A \\
00000004\_0 & 6.2 & 8.284 & 0.214 & 6.230 & 0.004 & 0.227 & 0.014 & 1p & A \\
00000004\_0 & 6.2 & 3.397 & 0.486 & 6.216 & 0.007 & 0.124 & 0.022 & 2p & A \\
00000004\_0 & 6.4 & 5.661 & 0.418 & 6.320 & 0.014 & 0.483 & 0.034 & 2p & --- \\
00000007\_0 & 6.2 & 0.750 & 0.036 & 6.300 & 0.007 & 0.290 & 0.018 & 1p & C \\
00000010\_0 & 6.2 & 3.772 & 0.263 & 6.256 & 0.013 & 0.376 & 0.036 & 1p & B \\
00000029\_0 & 6.2 & 2.375 & 0.283 & 6.225 & 0.010 & 0.159 & 0.025 & 1p & A \\
\enddata
\tablecomments{Table 2 is published in its entirety in the electronic edition of the {\it Astrophysical Journal}.  A portion is shown here for guidance regarding its form and content. {Label descriptions: IDEOS\_name is the name in the IDEOS catalog, z is the redshift, Band is the band identifier (6.0 as water ice band and 6.2 and 6.4 as PAH bands),  Amp is the fitted amplitude, e\_Amp is the error in fitted amplitude, $\lambda_c$ is the fitted central wavelength, e\_$\lambda_c$ is the error in fitted central wavelength, FWHM is the fitted full width at half maximum, e\_FWHM is the error in fitted FWHM, fit\_type is the type of fit (1p, 2p or ice), and Peeters\_class is the Peeters' class (A, B or C). }}
\end{deluxetable*}

\subsection{The band profile}

Taking into account the instrumental uncertainties, which are not included in the fit errors, the spectral resolution element of Spitzer/IRS is 0.10-0.05~$\mu$m. Those systematical uncertainties are expected to be more equally distributed in a large database in order to not affecting the overall results. {Moreover, \citet{hernan2016} and \citet{Spoon_2022} assessed uncertainties associated with flux calibration, slit losses, and stitching of spectral segments in IDEOS. They verified that segment-to-segment scaling uncertainties are generally small ($<$5\%), demonstrating that these instrumental effects are not significant when compared to the fitting errors.} To confirm how the instrumental uncertainties interfere with the results, more high resolution observations are needed,  reinforcing the importance of our analysis in the JWST era. 

In general, differences in PAH profiles can be attributed, for instance, to  local ISM physical conditions and of the PAH molecules' size, charge, geometry and heterogeneity \citep[e.g][]{Draine01, Draine07, Smith07, sales13}. Taking into account the CC vibration modes of the 5 -- 9~$\mu$m wavelengths, the peak positions and the profiles themselves can be highly variable, normally produced by ionized PAHs \citep[e.g.][]{Tielens08}. In particular, the 6.2~$\mu$m seems to be deeply influence by PANHs, including the position of the nitrogen in the rings. According to \citet{Vats2022}, PAHs with NH$_2$ and N inside the carbon structure show the band features characteristic to star-forming regions as well as reflection nebulae (Class A profiles), whereas PAHs with N at the periphery have similar spectra to Class B profiles, observed in planetary nebulae and post-AGB stars. However, \citet{Ricca2021} exclude PAHs with NH$_2$ peripheral groups in view of the absence of a strong NH stretch in the ISM spectra. Again,  addressing this issue is not what we propose here.

\subsubsection{Water ice absorption feature}

\begin{figure*}[ht!]
\includegraphics[width=\linewidth, keepaspectratio]{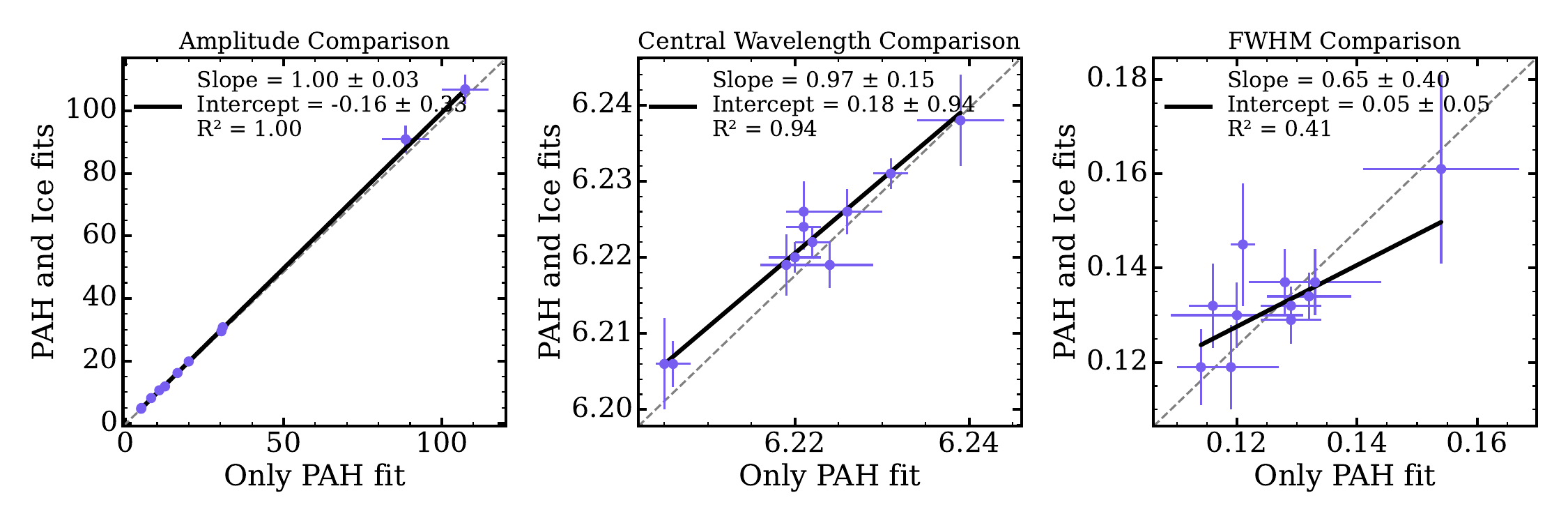}
\caption{Comparison of the 6.2~$\mu$m PAH fit results amplitude, central wavelength and FWHM considering just one PAH profile (x axis) and fitting together with the water ice feature (y axis). The gray dashed line shows a slope of 1 for comparison. The black line shows the linear regression for each parameter, with the slope,  interception {and R-squared (R$^2$)} values displayed {within the panels}.}
\label{fig:ice}
\end{figure*}

A total of eleven sources have the water ice absorption fitted, of which only two are SB and the others are mixed-dominated objects. To properly compare the results, we fitted a {linear} regression using the uncertainties as weights for the three 6.2~$\mu$m band profile parameters. As can be seen in Figure~\ref{fig:ice}, the amplitude and central wavelength are poorly affected by absorption. The FWHM, on the other hand, presented a slope of 0.65. In this sense, the Peeters' classification is not compromised by the fit of the water ice, but the PAH equivalent width (PAH$_{6.2}$EW) calculation must consider it, as performed by \citet{Spoon_2022}. For this reason, we will use previous PAH$_{6.2}$EW values obtained by IDEOS in our analysis. 

\subsubsection{A second PAH emission feature}

Unlike the SB sample, in which it was possible to identify several asymmetric profiles with a blue rise and a red tail,  AGN~1 sample showed more symmetric profiles without the blue rise and, in some cases, with a smoothed red tail. This tail could be just a characteristic of the anharmonic profile \citep[e.g.][]{Tielens08} and it is not possible to be well-fitted with the addition of another profile. As {a} matter of fact, ionized PAH would show very similar anharmonicity effects on this profile \citep{Mackie2022}. This second profile could be interesting for providing a measure of the importance of anharmonicity. \citet{Mackie2022} also show for CHOOPS mode in neutral PAHs that the wing is more pronounced when the internal excitation is higher. A similar effect is expected for cationic PAHs and the CC stretch mode.  In general, the presence of a more prominent second feature occurs at low-intensity sources.  In our sample, we avoided observations of lower signal-to-noise ratio  to not interfere with the final best-fit analysis. {We established a limit of S/N $\geq$ 10 at 6.6~$\mu$m, below which spectra could not be reliably fitted. The S/N values were extracted from IDEOS. Moreover, we focus on the most prominent features to avoid mistakenly fitting the intrinsic asymmetry of class A profiles as a second PAH component. While multiple Gaussian fits may improve the overall fit quality, they do not necessarily represent distinct PAH populations.}  

\begin{figure*}[ht!]
\includegraphics[width=\linewidth, keepaspectratio]{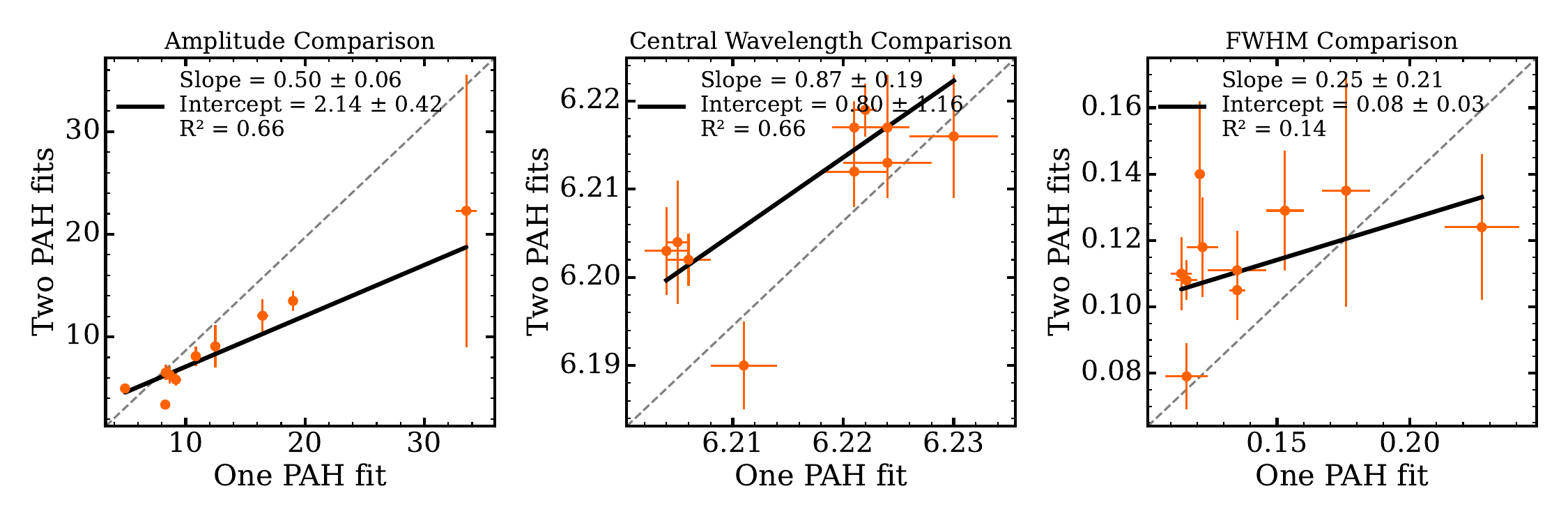}
\caption{Comparison of the 6.2~$\mu$m PAH fit results amplitude, central wavelength and FWHM considering just one PAH profile (x axis) and two PAH profiles (y axis). The gray dashed line shows a slope of 1 for comparison. The black line shows the linear regression for each parameter, with the slope, interception {and R-squared (R$^2$)} values displayed {within the panels}.}
\label{fig:2p}
\end{figure*}

The fit of the second emission can be done independently of the water ice absorption, because it is mainly the FWHM that is affected in the previous case. However, fitting two PAH profiles could change the results of the first band on a larger scale, especially for amplitude and FWHM values, as seen in Figure~\ref{fig:2p} with the weighted {linear} regression fits. Nevertheless, JWST observations of the Orion Bar revealed an extremely weak second feature on the wing of 6.2~$\mu$m band that does not interfere with the {strength's} wing determination \citep{Chown2024}. In our sample, ten sources have the most prominent second feature fitted, of which one is SB, another is AGN1 and the remaining, just as in the water ice scenario, are mixed-dominated objects. With these prerequisites, we obtained a maximum central wavelength of {6.35~$\mu$m}. 

{Although we do not have an example of this situation in our sample}, this second PAH emission can occur at higher wavelengths.  For instance, emissions at nearly 6.41~$\mu$m could be due to perylene-like structures in the PAHs, according to \citet{Candian14}. In the Reflection Nebula NGC7023, the observed 6.4~$\mu$m feature was attributed to the C$^+_{60}$ \citep{berne15}. At higher wavelengths (6.6 to 6.7~$\mu$m, approximately), it was suggested emissions from the planar graphene form of C$_{24}$ carbon cluster \citep[][and references therein]{Sadjadi2020}. Moreover, the presence of such PAH complex structures can be expected at hostile environments once the size of the formed molecules is found to roughly increase with increasing temperature up to 800~K, and to be correlated with the level of dehydrogenation \citep[e.g.][]{Hanine2020}. 

Recently, \citet{Xu2023} derived vibrational properties of fullerene species and created a database that includes their emission frequencies. According to their study, the following species present emissions that could account for the second feature observed in some of our galaxies, although they may have features at other wavelengths that exclude an important contribution in this band: C$_{70}$ and C$^{+}_{60}$ at 6.5~$\mu$m, C$_{60}$H$^{+}$ at 6.5, 6.6 and 6.8~$\mu$m, C$_{60}$O$^{+}$ at 6.5, 6.6 and 6.7~$\mu$m, C$_{60}$OH$^{+}$ at 6.5, 6.6, 6.7 and 6.8~$\mu$m, C$_{70}$H$^{+}$ at 6.5, 6.7 and 6.8~$\mu$m, and C$_{60}$V$^{+}$ at 6.5, 6.6, 6.7 and 6.8~$\mu$m. This also contributes to the suggestion that fullerene species can survive and are abundant in AGN environments.  We calculated the ratio between the 6.4 and 6.2~$\mu$m fluxes in order to analyze the importance of each feature in different environments. Unfortunately, we only have one AGN~1 source, in which the second feature presented a strong and significant contribution to the band. In the mixed and SB objects, the ratio varied between 0 and 2.6, as seen in Table~\ref{tab:ratio64}. Therefore,  high-resolution data are necessary to validate the existence of the potential second feature in astrophysical environments, together with more laboratory and modeling to identify its carriers and distinguish them from the 6.2~$\mu$m band asymmetry.

\begin{deluxetable*}{cccc} 
\tablewidth{0pt}
\tablecaption{Flux ratio between 6.4 and 6.2~$\mu$m features.\label{tab:ratio64}}
\tablehead{
\colhead{IDEOS\_name} & \colhead{grouped\_IDEOS\_class} & \colhead{Flux\_Ratio} & \colhead{e\_Flux\_Ratio} \\  
\colhead{} & \colhead{} & \colhead{1e-17 W/m$^2$} & \colhead{1e-17 W/m$^2$}
}
\startdata
00000004\_0 & AGN1 & 6.28 & 1.57 \\
10103552\_0 & mixed & 1.71 & 0.36 \\
10104320\_0 & SB & 1.06 & 0.27 \\
10107904\_0 & mixed & 2.54 & 0.54 \\
10111232\_0 & mixed & 1.09 & 0.30 \\
10869504\_0 & mixed & 0.79 & 0.44 \\
18510848\_0 & mixed & 1.43 & 1.15 \\
25186816\_0 & mixed & 0.28 & 0.07 \\
4966144\_0 & mixed & 1.33 & 0.44 \\
4968448\_0 & mixed & 0.88 & 0.35
\enddata
\end{deluxetable*}

\subsection{The Peeters' classes}

After fitting of the 6.2~$\mu$m band, it is possible to perform the distribution of the profiles into the Peeters' classes, according to their central wavelengths. When water ice absorption is {presented}, we use the fit results from "ice fit\_type". The general percentages for each class are shown in Table~\ref{tab:results-classes} and the same classification is performed for the \citet{Canelo18} in Appendix~\ref{sec:ap}. As can be seen, class A is predominant in the three main types of galaxies, especially in the SB and mixed objects. It also consists of 80\% of our entire sample, followed by classes B and C, respectively. In fact, class C has not appeared in SB objects at all. These results are expected because class A is more connected with young stellar environments, which are characteristic of starburst galaxies and are present in mixed-dominated ones. As we preferred spectra with stronger signal-to-noise ratios, it is {expected to obtain} intense PAH emission from those environments.  Nevertheless, it is important to point out the elevated percentage of class A galaxies, that can indicate molecular environments with high {chemical} complexity, including PAHN molecules. 

In the meantime, class C profiles appear normally in evolved and dusty objects, which could be correlated with the central torus that protects galaxies from the SBMH radiation as well as emission from the ISM of the galactic disks and regions external to AGNs. {In fact, \citet{Florido2015} have shown that the nuclear regions of galaxies are overabundant in nitrogen. Class C appears more frequently in the AGN~1 objects, which could be related to the dusty material of those galaxies. In the context of inside-out growth of discs, our findings suggest that central regions of massive galaxies today have evolved to an equilibrium metallicity, while the nitrogen abundance continues to increase as a consequence of delayed secondary nucleosynthetic production \citep[e.g.][]{Belfiore2017}. }The Peeters classification of AGN~1 object is not as uneven as in the other sources, which clearly indicates the difference between SB and AGN-dominated galaxies.  In order to ensure a non-biased study of AGN~1 and ~2 objects, data with higher {spectral and spatial} resolution are needed.

\begin{deluxetable*}{ccccc}
\tablewidth{0pt}
\tablecaption{Distribution of the galaxies into the Peeters' classes for the three type of galaxies.
	\label{tab:results-classes}}
\tablehead{
\colhead{Sample} & \colhead{Class A (\%)} & \colhead{Class B (\%)} & \colhead{Class C (\%)} & \colhead{Total (\%)}
}
\startdata
SB & 13.7 & 0.6 & --- & 14.3 \\
AGN~1 & 22.3 & 9.7 & 2.3 & 34.3 \\
Mixed & 44.0 & 6.8 & 0.6 & 51.4 \\
\hline
Total & 80 & 17.1 & 2.9 & 100 \\
\enddata
\end{deluxetable*}

\begin{figure*}[ht!]
\includegraphics[width=\linewidth, keepaspectratio]{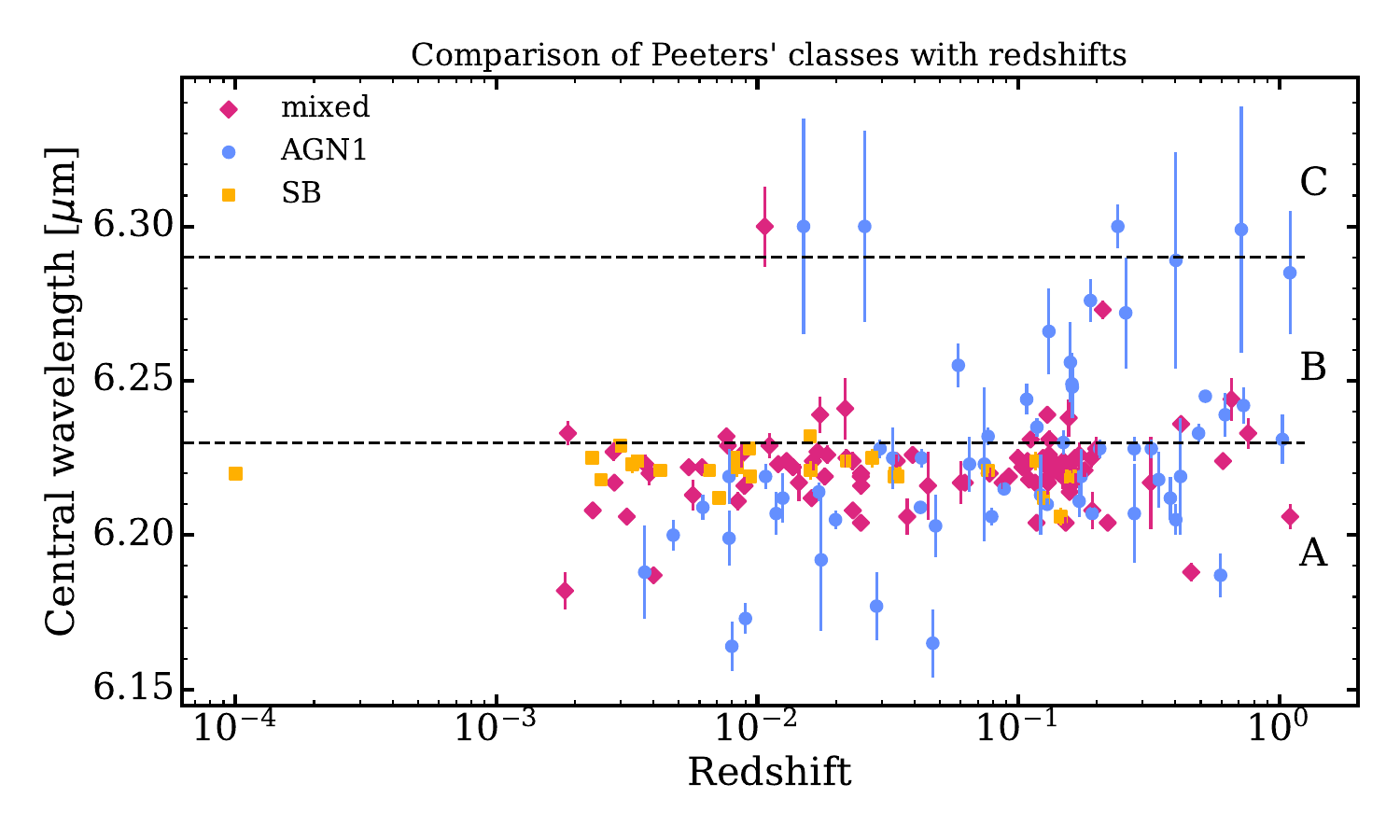}
\caption{Distribution of the 6.2~$\mu$m bands for the  SB, AGN~1 and mixed-dominated objects according to the galaxies' redshift. The dashed lines are the limits among the Peeters' classes, indicated also by A, B or C letter.}
\label{fig:z}
\end{figure*}

To better understand the differences between the SB, AGN~1 and mixed-dominated objects, focusing on the 6.2~$\mu$m band,  Fig.~\ref{fig:z} shows the Peeters' classification along the redshift of the sources. \citep{Canelo18} have noticed that class B and C seem to be more abundant at higher redshifts, as can be seen in {Table-\ref{tab:z}.} Although this may be just an observational bias, it could imply on a PAH evolutionary timescale \citep[similar to the stellar lifecycle and aromatic evolution of][]{Shannon19}. At higher redshifted galaxies, the spectra may include emission from star-forming regions due to the angular resolution of the Spitzer telescope. This indicates that, despite the contamination of SB emission that may classify that galaxy as a class A source, we could still see more class B and C objects. 
{Given that AGN tori typically extend to scales of $\sim$1 kpc from the nucleus, our observations do sample the full spatial extent of the circumnuclear region at higher redshifts, including potential contributions from star-forming regions. To evaluate the robustness of our classification against such contamination, we note that \citet{Canelo21b} analyzed extinction effects in the Seyfert galaxy Mrk~52 using the extinction curve from \citet{Hensley2020}. By simulating spectra with dust contamination from silicates at 9.8~$\mu$m, they found that Peeters' classification at 6.2~$\mu$m was robust against extinction effects. However, PAH bands at longer wavelengths were more severely compromised, which could affect classification at higher redshifts. Although a comprehensive analysis of these longer-wavelength bands is deferred to a future study, we acknowledge that } JWST has the potential to better address this issue \citep[e.g.][]{Stiavelli09}.

\begin{deluxetable*}{cccc}
\tablewidth{0pt}
\tablecaption{Distribution of the Peeters' classes along the galaxies redshifts (z).
	\label{tab:z}}
\tablehead{
\colhead{Class} & \colhead{z $\leq 10^{-2}$} & \colhead{$10^{-2} <$ z $\leq 10^{-1}$} & \colhead{z $> 10^{-1}$} 
}
\startdata
A & 25.0~\% & 36.4~\% & 38.6~\% \\
B & 6.7~\% & 16.7~\% & 76.6~\% \\
C & 0.0~\% & 60.0~\% & 40.0~\% \\
\enddata
\end{deluxetable*}

\subsection{Advanced analysis}

\begin{figure*}[ht!]
\centering
\includegraphics[width=\linewidth, keepaspectratio]{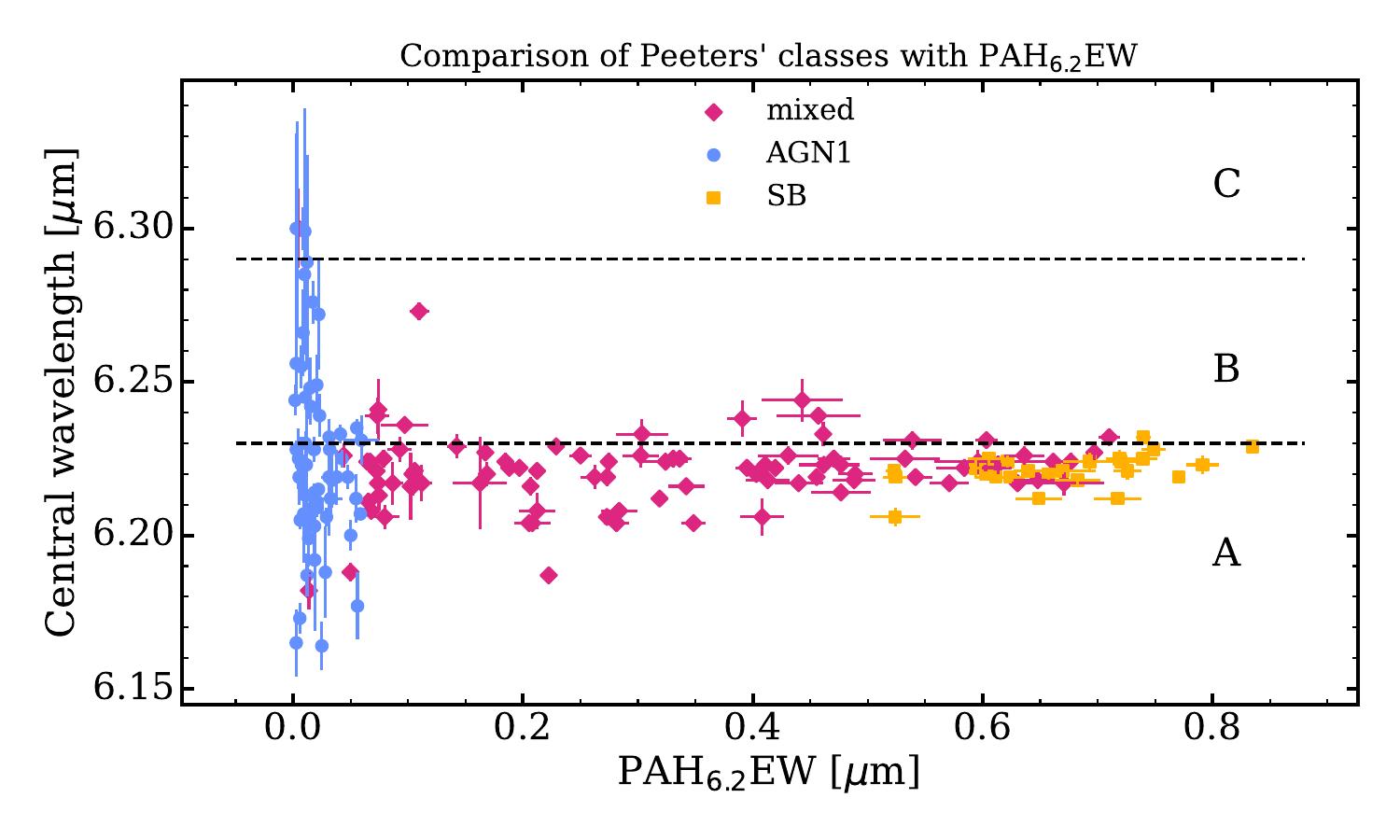}
\includegraphics[width=\linewidth, keepaspectratio]{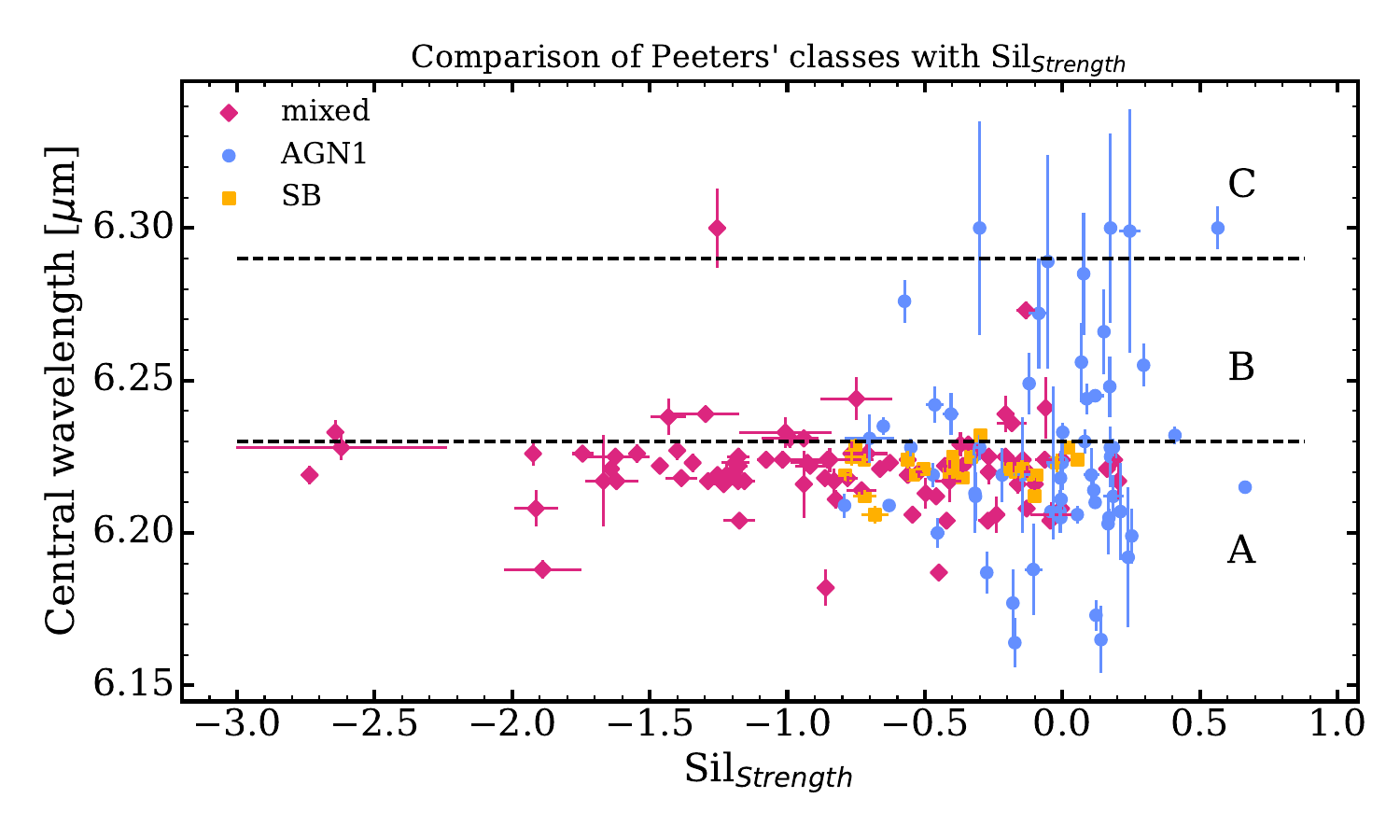}
\caption{Comparison between the central wavelengths and the equivalent width and silicate strength.  The dashed lines are the limits among the Peeters' classes, indicated also by A, B or C letter.}
\label{fig:ew-sil}
\end{figure*}

We decide to explore the IDEOS database and verify if there is any correlation of the 6.2~$\mu$m PAH band central wavelength and, consequently, the Peteers' classes, with its equivalent width and the silicate strength at 9.8~$\mu$m (Sil$_{Strength}$). {The Sil$_{Strength}$ is the relation between the flux density of the peak at 9.8~$\mu$m and the flux density of the continuum at the same wavelength \citep{Spoon_2022}. }As shown in Figure~\ref{fig:ew-sil}, the PAH$_{6.2}$EW is well connected with the type of objects, once the MidIRClasses used its value to classify the sources. There is no correlation between the central wavelengths of AGN~1 and the Peeters' classification.  However,  almost {all} SB and mixed galaxies present a clear tendency to  have an intermediate Peeters´ class A or B. Considering Sil$_{Strength}$, class B and C sources tend to present higher values, while class A objects are spread along the range.

\section{Conclusions}
\label{sec:conclusion}

{This work presents a comprehensive analysis of PAH 6.2~$\mu$m profiles in 175 galaxies from the IDEOS database using Peeters' classification scheme. The main findings are:}

\begin{itemize}
    \item {Class A dominance:} Class A objects are predominant in 80\% of the entire sample, with particularly high fractions in starburst and mixed-dominated galaxies, suggesting the presence of PANHs in environments with active star formation and/or nuclear activity.
    
    \item {Water ice contamination:} Water ice absorption affects the FWHM and equivalent width of the 6.2~$\mu$m band but does not compromise the Peeters' classification itself.
    
    \item {Secondary PAH feature:} A second spectral feature at $\sim$6.35~$\mu$m is detected in our sample and may be associated with perylene-like PAH structures, C$^+_{60}$, or other fullerene species. High-resolution observations are needed to validate this interpretation.
    
    \item {Redshift trends:} The distribution of Peeters' classes with redshift follows the trend observed in previous studies, though observational selection bias toward luminous galaxies cannot be excluded.
    
    \item {Silicate strength correlation:} Greater silicate absorption strengths are associated with classes B and C, while PAH equivalent width shows no correlation with classification.
\end{itemize}

{Future perspectives: To advance our understanding of PAH properties across cosmic environments, we recommend: (1) extending the analysis to longer-wavelength PAH bands (7.7, 8.6, and 11.2~$\mu$m); (2) including AGN~2 objects and higher-redshift sources; (3) incorporating chemical evolution models that account for metallicity, star formation history, and molecular cloud properties; (4) combining computational predictions with laboratory and observational measurements; and (5) utilizing JWST observations to achieve the spatial resolution and sensitivity necessary to study PAH evolution and constrain the timescales of PAH processing in different galactic environments.}

\begin{acknowledgments}
CMC acknowledges the support of Coordena\c{c}\~ao de Aperfei\c{c}oamento de Pessoal de N\'ivel Superior - Brasil (CAPES) - Finance Code 001 - Brazil, process number 88887.967338/2024-00.
\end{acknowledgments}

\appendix

\section{Comparison with previous analysis}
\label{sec:ap}

In order to evaluate the importance of accurate data on Peeters' classification, we focused on the comparison of the central wavelengths from \citet{Canelo18} and ours. As seen in the previous sections, the central wavelength is still the main component to determine the Peeters' class of a profile. A total of 75 objects are in both studies, and we also distributed the classes according to the grouped IDEOS classification, as shown in Table~\ref{tab:app}. As can be seen, the same trend presented here has also appeared in the previous work.

\begin{deluxetable*}{ccccc}
\tablewidth{0pt}
\tablecaption{Distribution of the \citet{Canelo18} galaxies into the Peeters' classes for the three type of galaxies
	\label{tab:app}}
\tablehead{
\colhead{Sample} & \colhead{Class A (\%)} & \colhead{Class B (\%)} & \colhead{Class C (\%)} & \colhead{Total (\%)}
}
\startdata
SB & 28 & 1 & --- & 29 \\
AGN~1 & 1 & --- & --- & 1 \\
Mixed & 62 & 8 & --- & 70 \\
\hline
Total & 91 & 9 & --- & 100 \\
\enddata
\end{deluxetable*}

\begin{figure}[ht!]
\includegraphics[width=\linewidth, keepaspectratio]{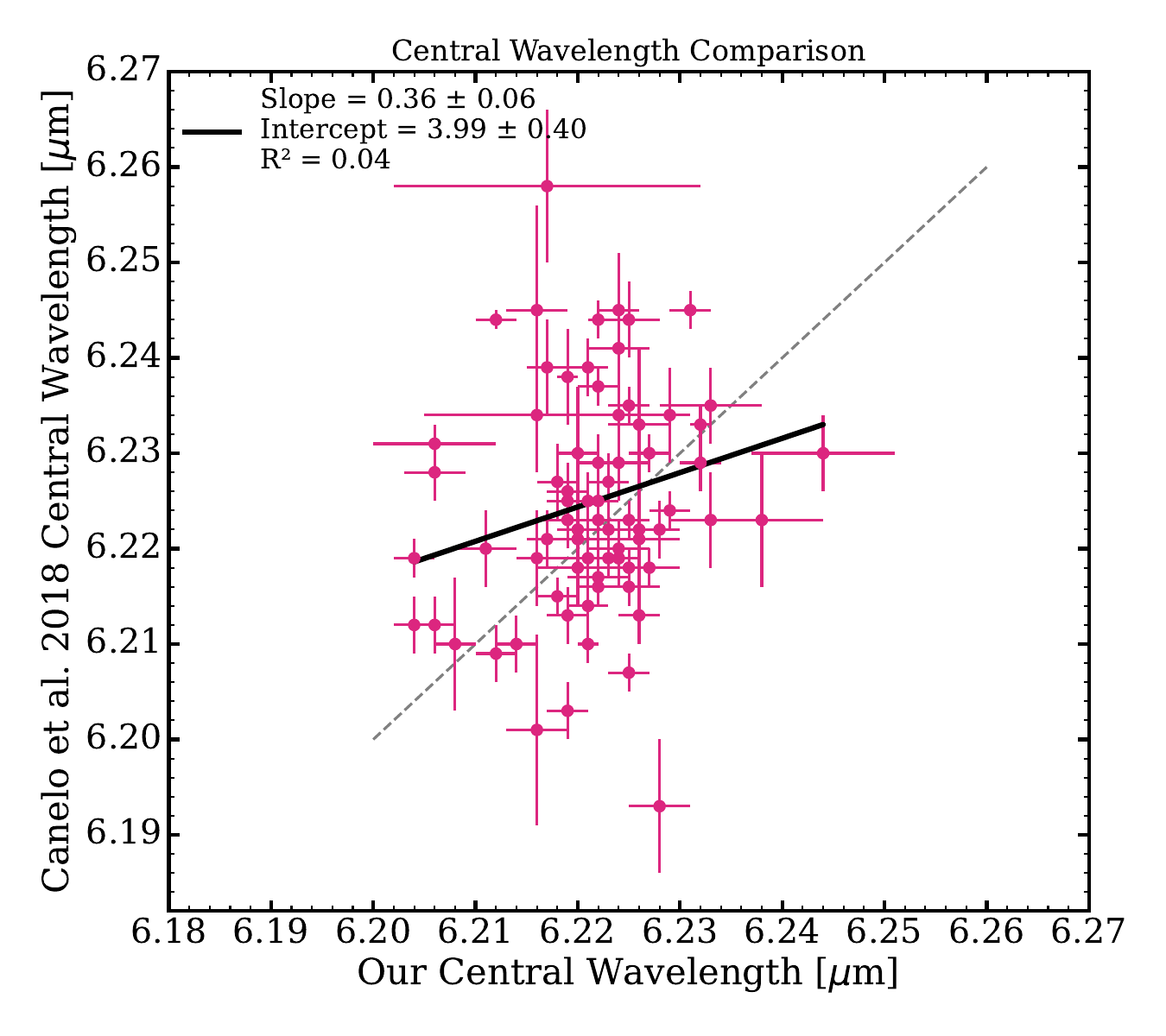}
\caption{Comparison of the 6.2~$\mu$m PAH fit results central wavelength from \citet{Canelo18} and our analysis. The gray dashed line shows a slope of 1 for comparison. The black line shows the linear regression, with the slope, interception {and R-squared (R$^2$)} values displayed {within the panels}.}
\label{fig:2018}
\end{figure}

{We calculated the Pearson correlation coefficient between the two samples to assess their relationship. The analysis revealed an extremely weak and statistically non-significant correlation (r = 0.158, p = 0.176). Given this negligible correlation, we proceeded with a weighted linear regression analysis, which yielded a slope of 0.36, further confirming the poor linear relationship between the samples (Figure~\ref{fig:2018}}). From a statistical point of view, the distribution of the results would not interfere with the classification in a large scale, but the need for high resolution data is evident. 

\section{Fitting the continuum emission with \textsc{PAHFIT-MCMC}}
\label{sec:ap2}

\begin{figure}[h]
\includegraphics[width=\linewidth, keepaspectratio]{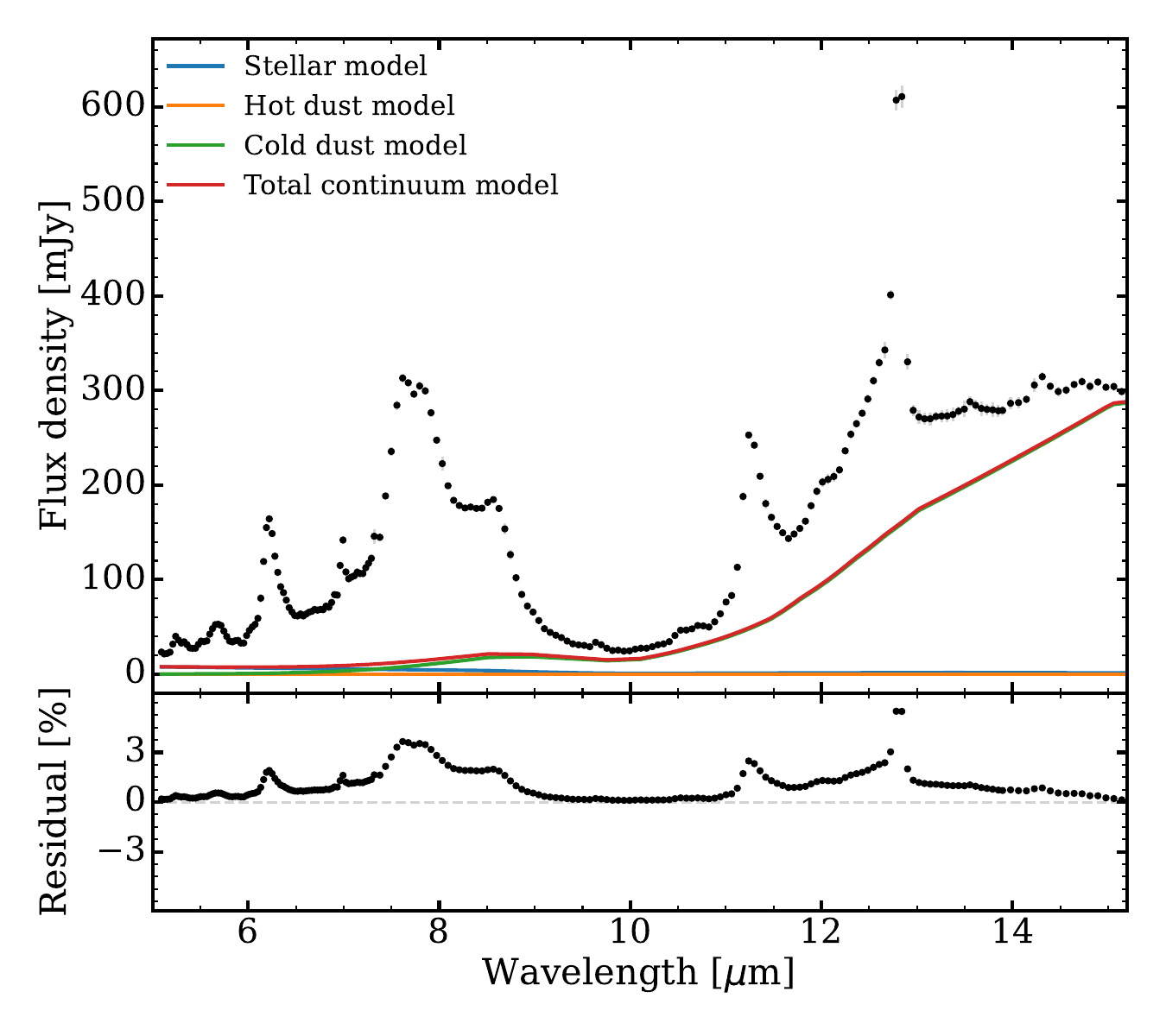}
\caption{Continuum emission fitted with \textsc{PAHFIT-MCMC} for galaxy 9072896\_0. The total emission and the contributions from the stellar, cold dust and hot dust components are also shown. }
\label{fig:pahfit}
\end{figure}

Figure~\ref{fig:pahfit} shows the continuum contributions decomposed by \textsc{PAHFIT-MCMC} for the galaxy 9072896\_0. As can be seen, the PAH plateaus are not well reproduced by this tool. To address this limitation, Drude profiles are adopted to fit the PAH bands, as their broader wings are better suited to account for the extended plateau emission underlying the individual features.

\bibliography{bib}{}
\bibliographystyle{aasjournal}



\end{document}